\documentclass[a4paper,11pt]{article}
\usepackage{jinstpub} 
\usepackage{lineno}
\usepackage{placeins}
\usepackage{graphicx}
\usepackage{subcaption}

\title{\boldmath Study on the shielding efficiency of water, HDPE,  and boron-loaded HDPE for neutron background of plastic scintillator neutrino detector}

\author[a,1]{D.X. Lu,}
\author[a]{Y.H. Liu,}
\author[a]{X.S. Zhang,}
\author[a,*]{F.P. An,}
\author[c,*]{G. Luo,}
\author[a,b,*]{W. Wang,}
\affiliation[a]{School of Physics, Sun Yat-sen University,\\
No.135 West Xingang Road, 510275 Guangzhou, China}
\affiliation[b]{ Institut Franco-Chinois de l'Énergie Nucléaire , Sun Yat-sen University,\\
No. 2 Daxue Road,519082 Zhuhai, China}
\affiliation[c]{School of Science, Sun Yat-sen University,\\ No. 66 Gongchang Road, 518107 Shenzhen, China}

\emailAdd{anfp@mail.sysu.edu.cn}
\emailAdd{luog7@mail2.sysu.edu.cn}
\emailAdd{wangw223@mail.sysu.edu.cn}

\abstract{Surface-level reactor antineutrino experiments usually have substantial cosmic-ray-induced neutron backgrounds, particularly with shallow overburden. The Array of Lattice for Anti-neutrino Reactor Monitoring (ALARM) is a plastic scintillator–based experiment designed for reactor power monitoring. It will be deployed about 44 m from the core of a reactor at the Taishan Nuclear Power Plant. Placed at a depth of 9.6 meters below the surface, cosmic‑ray‑induced fast neutrons constitute a significant background, making an effective neutron shielding system essential for the experiment. For the shielding design of ALARM, we tested the shielding performance of three materials—water, HDPE, and 40\% boron-doped HDPE (BHDPE)—against both fast and thermal neutrons. A thermal neutron detector composed of an EJ426 scintillator setup was first used to measure the shielding efficiency of these materials at various thicknesses using neutrons from an Am-Be source. A 30-cm thickness of BHDPE achieved a shielding efficiency exceeding 95\% for both fast and thermal neutrons. Monte Carlo simulations of the EJ426 setup yielded results consistent with the experimental data. Simulation results for the shielding performance of the full ALARM shielding assembly are also presented.}


\begin{document}
\maketitle
\flushbottom

\section{Introduction}
\label{sec:intro}

Nuclear reactors are considered to be an intense sources of electron antineutrinos. Since neutrinos can penetrate the reactor shielding, neutrino detectors can be placed outside the reactor buildings, offering the advantages of being non-invasive and enabling remote monitoring. These detectable neutrinos can be utilized in various reactor applications, such as monitoring the power of reactors, measuring the fissile content of reactors, etc~\cite{akindele2021nu}. These applications require a relatively small-volume antineutrino detector with adequate efficiency, which demands compact shielding.

Due to the extremely rare interactions of antineutrinos, numerous backgrounds can interfere with neutrino detection, reducing the sensitivity of experiments. The most common backgrounds come from internal radioactivity, ambient neutrons and gammas~\cite{chen2021radiogenic},  cosmic muon induced background~\cite{PhysRevD.97.052009, chen2021radiogenic,Luo_2025}, etc. Among these neutron backgrounds are the most important ones since the neutrino candidate is usually tagged by firstly finding a neutron event.  In this study we focus on the neutrons come from outside of the detector, including the ambient neutrons and the cosmic muon induced neutrons.

There are typically two approaches to dealing with neutron backgrounds. One is passive shielding using thick shielding layers, and the other is to use active tagging to veto neutron-related events. However, when the detector is placed closer to the reactor or or when the overburden is shallow, active tagging alone cannot effectively shield against the substantial background from the reactor and cosmic rays, and shielding layers are still required. Hydrogen-rich materials are often used as shielding layers for reactor detectors. For example, the PROSPECT experiment~\cite{ashenfelter2019prospect} and the Taishan Anti-neutrino Observatory (TAO) experiment~\cite{li2022detector} both used High-Density Polyethylene (HDPE) and water to shield against neutrons.

The Array of Lattice for Anti-neutrino Reactor Monitoring (ALARM) is a compact reactor neutrino experiment designed to monitoring the nuclear reactor power via antineutrino detection, aiming to achieve a  monitoring precision of 5\%–10\%, which will be deployed in the experimental hall at the Taishan Nuclear Power Plant in Guangdong Province, China. The dominant background in this experiment arises from fast neutrons generated by cosmic‑ray muons interacting inside and outside the detector, and from the ambient neutrons. To mitigate neutron backgrounds, the experiment plans to employ a passive shielding layer, with careful selection of shielding materials and their thicknesses to identify the most efficient and cost‑effective solution. 

To test the neutron-shielding performance with different materials and thicknesses, we employed an Am-Be neutron source and used a simplified thermal neutron detector consisting of a single-layer EJ426 scintillator~\cite{eljen426} coupled to a photomultiplier tube (PMT). The shielding materials evaluated included HDPE, boron-doped HDPE (BHDPE), and water. We then simulate the overall shielding effectiveness of the detector materials in the neutron environment of the Taishan experimental hall, which can serve as a reference for future reactor neutrino detector shielding designs. We briefly describe the design of the ALARM detector in Section~\ref{sec:design}. Shielding considerations regarding the neutron background, along with the selection and combination of shielding materials, are presented in section~\ref{sec:shielding}. Sections~\ref{sec:simulations} and~\ref{sec:results} present the simulation and experimental studies, respectively, on the shielding performance of the single‑layer EJ426 setup. Section~\ref{sec:alarmresults} describes the shielding simulation for the full ALARM detector, and conclude with a summary.

\section{ALARM detector design}
\label{sec:design}

\begin{figure}[htbp]
\centering
\includegraphics[width=.7\textwidth]{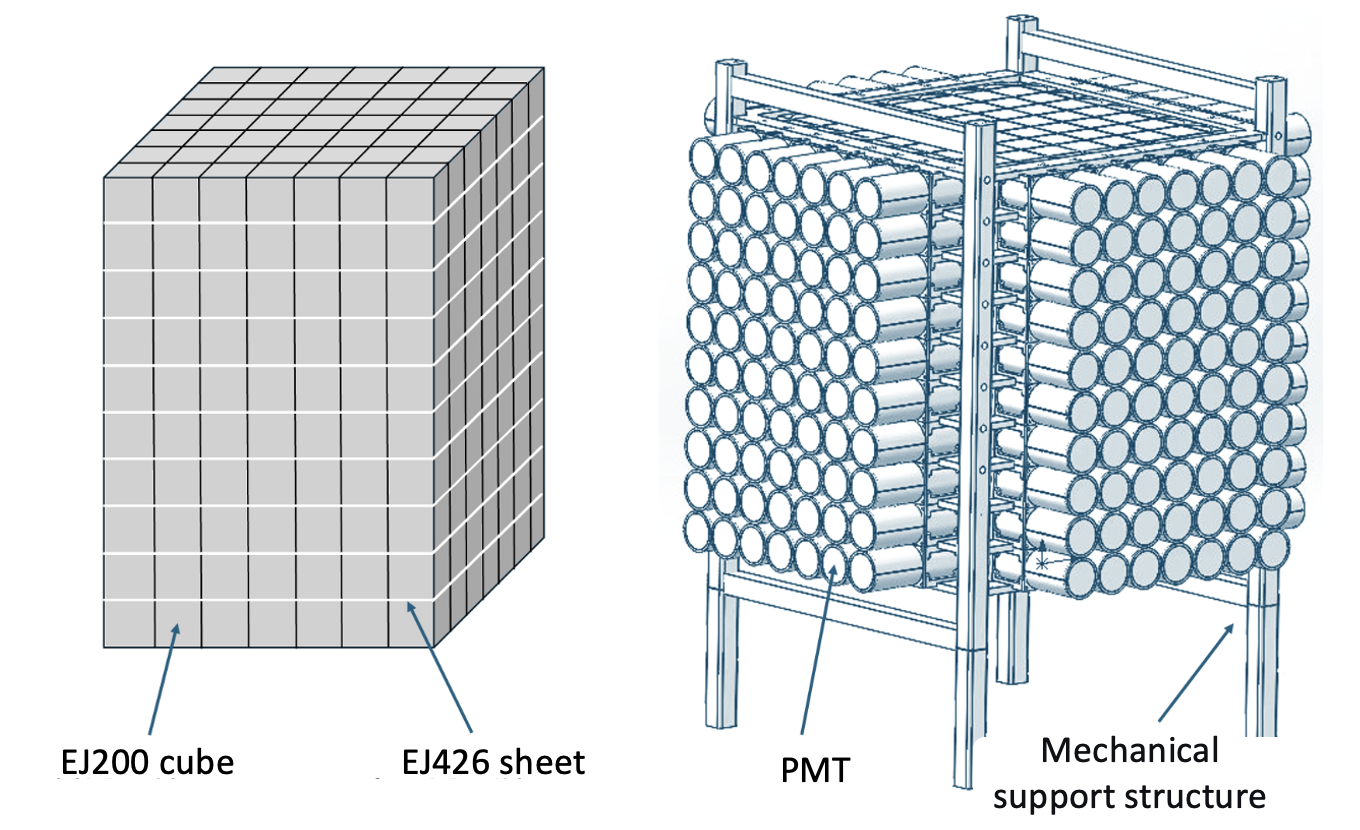}
\caption{Schematic view of the ALARM detector. The right panel shows the entire detector assembly. The left panel depicts the detector array, which consists of 10 layers, each composed of a 7×7 matrix of EJ200 scintillator cubes. EJ426 scintillator sheets are placed between consecutive layers.\label{fig:alarm_detector}}
\end{figure}

The ALARM detector will be installed in an underground hall approximately 44 meters from the reactor core at Taishan nuclear power plant. The hall is located at a depth of about 9.6 meters below the surface. The detector body consists of EJ200~\cite{eljen200} and EJ426 scintillators. A schematic diagram of the main detector is shown in figure~\ref{fig:alarm_detector}. Each layer of the detector is composed of a tightly packed 7×7 array of EJ200 plastic scintillator cubes. Light produced inside the cubes is guided via total internal reflection along the cube rows and columns. This light is detected by PMTs at the ends of the rows and columns, and can be used to reconstruct the position of an event within the cube located at the intersection of the active row and column. Between consecutive EJ200 layers, lithium-6 ($^{6}\text{Li}$) doped zinc sulfide (ZnS) EJ426 scintillator sheets (0.5 mm) are inserted, the optical isolation provided by these sheets effectively prevents light leakage between adjacent scintillator cube layers. A total of 11 such neutron-sensitive sheets are used in the detector. A 30-cm-thick layer of BHDPE is planned to surround the detector body to shield against environmental neutrons. Outside the neutron shielding, a 5-cm-thick lead brick layer will be installed to attenuate ambient gamma rays.

In this reactor neutrino detector, electron antineutrinos are observed through the inverse beta decay (IBD) reaction~\cite{haghighat2020observation}, producing a positron and a neutron:
\begin{equation*}
\bar{\mathrm{\nu}}_e + \mathrm{p} \rightarrow \mathrm{e^+} + \mathrm{n}
\end{equation*}

The positron deposits its kinetic energy in scintillator and releases energy in the form of gamma photons through processes such as ionization and annihilation, producing a prompt signal. Meanwhile, the neutron is thermalized and then captured by a lithium nucleus in EJ426, inducing the nuclear reaction $^{6}\text{Li}$(n,$\alpha$)$^{3}\text{H}$~\cite{PSD_and_exploration}:
\begin{equation*}
^{6}\mathrm{Li} + \mathrm{n} \rightarrow {}^{3}\mathrm{H} + {}^{4}\mathrm{He} + 4.78\ \mathrm{MeV}
\end{equation*}

The resulting triton and alpha particle then interact with ZnS:Ag, yielding a delayed signal with a more localized energy deposition. In this detector, the IBD interaction is identified through the temporal and spatial correlation between the prompt positron signal and the delayed neutron signal.

\FloatBarrier

\section{Shielding considerations}
\label{sec:shielding}

\subsection{Neutron background}
We categorize the neutron backgrounds into three types: fast neutrons, double neutrons, and others. Fast neutrons undergo elastic collisions with protons (hydrogen nuclei) in the scintillator, transferring part of their energy to the protons and producing recoil protons that generate a prompt signal. The neutrons are then thermalized and subsequently captured, producing a delayed signal. This prompt-delayed signal coincidence mimics the signature of an IBD event, thereby constituting a significant fast neutron background. Multiple neutrons produced by the same muon may be captured by multiple nuclei in the detector, creating prompt-delayed signal pairs, thereby forming the double neutron background. Others are backgrounds formed by accidental coincidences between some unrelated prompt-like and delayed-like events, with no correlation in space or time.

For the ALARM experiment, the majority of the neutron background is the fast neutron background, in which a neutron-proton collision first produces a prompt signal from the recoiling proton, followed by a delayed signal from the neutron. so the fast neutron background typically exhibits the temporal correlation of an IBD event, but with greater average spatial separation due to the greater energy and velocity of the initial neutron.

\subsection{Shielding materials}
Neutron shielding involves two key processes: moderation and absorption. Fast neutrons are first slowed down to thermal energies through interactions with the shielding material, primarily via inelastic and elastic scattering~\cite{introduction_shielding}. Subsequently, the thermalized neutrons are captured by the material, completing the shielding process. Due to the conservation law of energy and momentum, the energy transferred in a single elastic collision is maximized when the target nucleus has a mass comparable to that of the neutron. The hydrogen nucleus (proton), having nearly the same mass as the neutron, is therefore the most efficient moderator. This explains why hydrogen-rich materials are excellent candidates for fast neutron moderation.

Water is a common neutron shielding material, containing a large number of hydrogen atoms (H) with a hydrogen content of approximately 11.2\%~\cite{pubchem_water}. When a neutron undergoes an elastic collision with a hydrogen nucleus, the neutron transfers most of its energy to the hydrogen nucleus, thereby losing energy and being moderated into a thermal neutron. The thermalized neutrons have lower energy and are more easily captured by hydrogen nuclei in the water.

HDPE is also a hydrogen-rich material, with a hydrogen content of up to approximately 14.3\%~\cite{pubchem_hdpe}, making it highly effective for moderating fast neutrons. Adding boron elements such as boric acid or boron carbide to HDPE — referred to as boron-doped HDPE (BHDPE) — can further enhance its shielding performance. The $^{10}\text{B}$ isotope has a large thermal neutron absorption cross-section, reaching up to 767 barns~\cite{B_containing_nanomaterials}, allowing it to effectively absorb thermalized neutrons and generate harmless alpha particles.

This research will subsequently employ three materials—water, HDPE, and BHDPE with a boron mass fraction of 40\%—for a comparative study of their shielding performance.

\section{Shielding Simulation for Single-Layer EJ426 detector}
\label{sec:simulations}
A Geant4~\cite{Geant4:2003}-based EJ426 Single-Layer model has been developed for the calculations of the shielding performance of the proposed neutron shields. The model includes the geometry of the experimental setup, the properties of materials employed, and the characteristics of the neutron source.

\subsection{EJ426 physics model}
\begin{figure}[htbp]
\centering
\includegraphics[width=.8\textwidth]{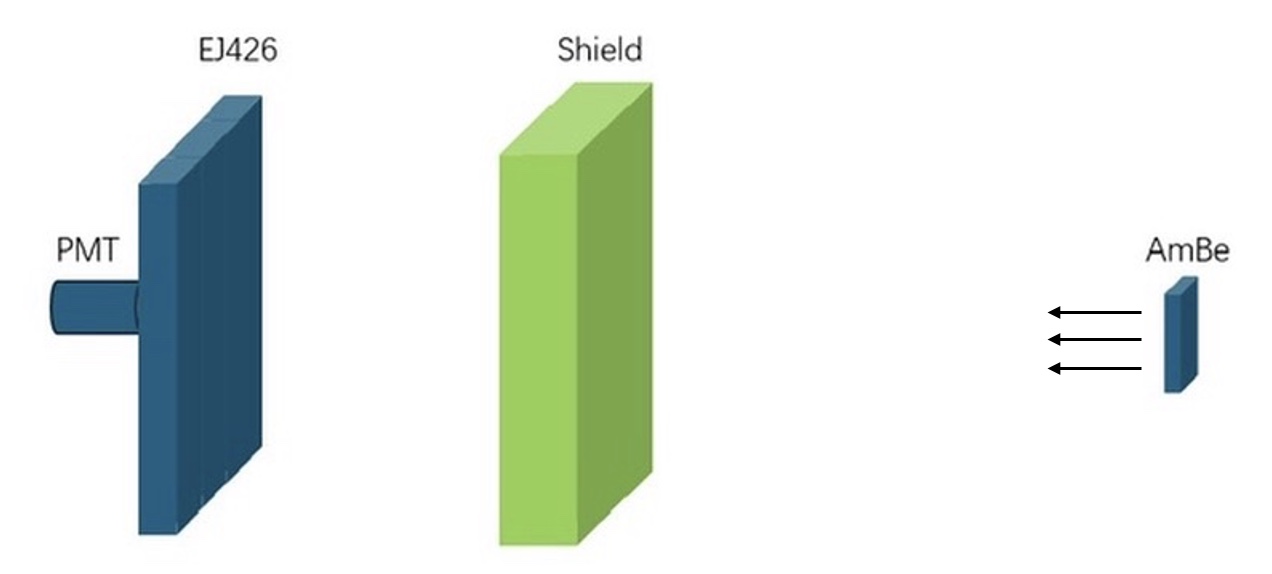}
\caption{General 3D view of the Single-Layer EJ426 model geometry.\label{fig:simulation setup}}
\end{figure}

As shown in figure~\ref{fig:simulation setup}, the Monte Carlo simulation uses the following configuration. The EJ426 scintillator, with the dimension of 420 mm × 420 mm × 0.5 mm, is oriented vertically and coupled to the PMT at its geometric center. The neutron source is modeled as a vertical square plane 50 mm × 50 mm, whose center is aligned with that of the EJ426 sheet and positioned 31 cm away. Neutrons are emitted uniformly in the horizontal direction toward the scintillator. The energy distribution of the source is based on the measured Am-Be neutron spectrum~\cite{iso8529_1}, which is imported into Geant4 with the General Particle Source (GPS) method. Between the detector and the source, a square shield of water, HDPE, or BHDPE—420 mm × 420 mm in lateral dimensions and 5–30 cm in thickness—is positioned with its center aligned with that of the EJ426 as well.

In each of the Monte Carlo runs, the program first tracks the transport and interactions of incoming neutrons within the shield, then records the number of initial incident neutrons that reach the EJ426 scintillator after passing through the shield. A fraction of these neutrons have their energy reduced to the thermal neutron range ($\boldsymbol{E}$ < 5 eV), where they undergo capture reactions with lithium-6 nuclei, producing secondary triton and alpha particles. We determine the number of captured thermal neutrons by recording these secondary particles.

\FloatBarrier

\subsection{EJ426 analysis results}
The main objective of the EJ426 simulation analysis is to predict the shielding effectiveness of various thicknesses of shielding materials (water, HDPE, and BHDPE) against both fast neutrons and thermal neutrons from the Am-Be source.

\begin{figure}[htbp]
\centering
\includegraphics[width=.7\textwidth]{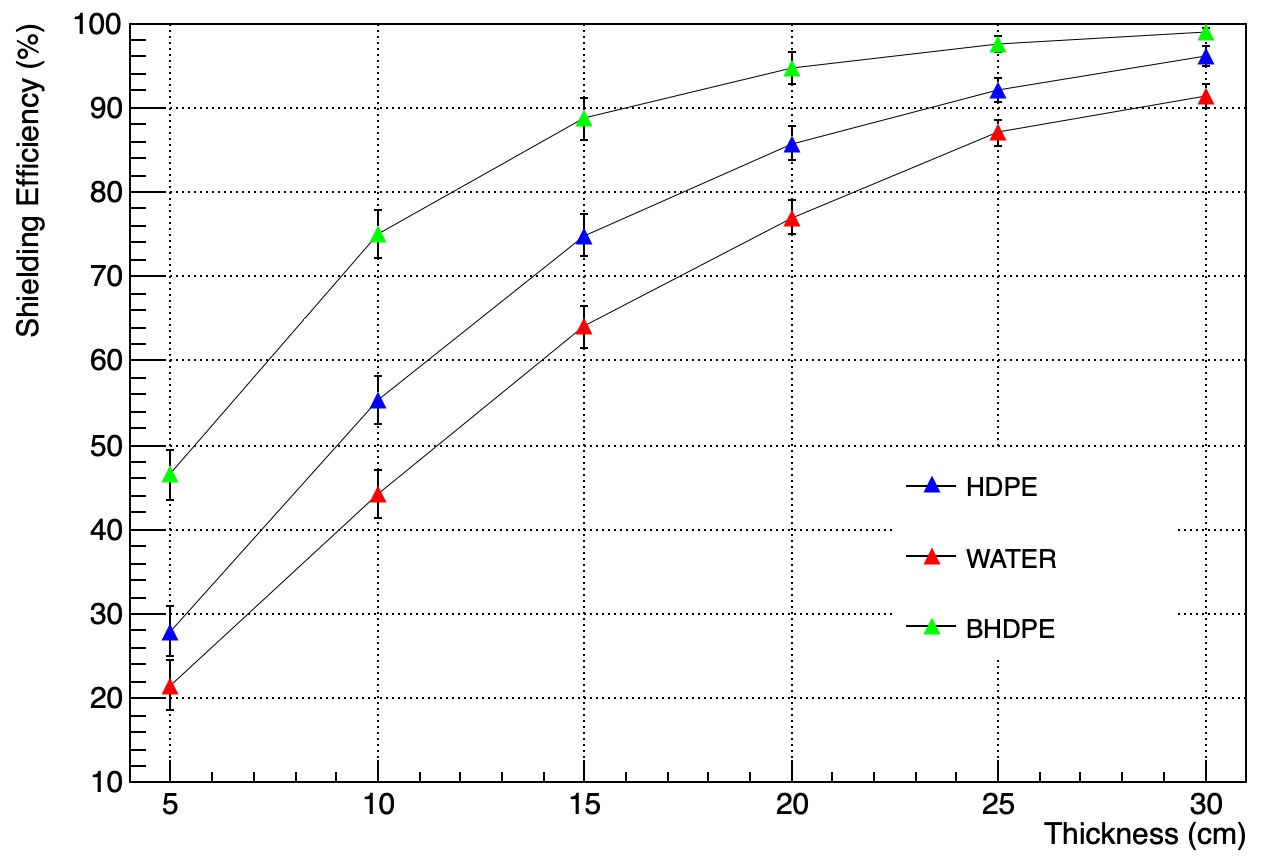}
\caption{Fast neutron shielding efficiency simulated with the EJ426 model for different materials.\label{fig:simulation all}}
\end{figure}

To quantitatively assess the shielding performance of a material against fast neutrons, figure~\ref{fig:simulation all} presents the fast neutron shielding efficiency, defined as
\[
R_{\text{block}} = \frac{N_{\text{blocked}}}{N_{\text{total}}}
\]

where \( N_{\text{blocked}} \) is the number of fast neutrons blocked by the shielding material and \( N_{\text{total}} \) is the total number of fast neutrons emitted from the source. The calculations indicate that for all material types, the increasing of shielding efficiency initially increases rapidly with increasing thickness, followed by a gradual reduction in the increasing rate. The shielding efficiency saturates at thicknesses of 25–30 cm, with the shielding efficiency exceeding 90\% for HDPE/BHDPE, and 85\% for water. Overall, at equivalent thicknesses, HDPE demonstrates a shielding efficiency up to 10\% higher than that of water, while BHDPE exhibits up to 20\% greater shielding effectiveness relative to pure HDPE.

After passing through the shielding material, some of the thermal neutrons reaching the EJ426 are captured by lithium nuclei and recorded by the oscilloscope. The number of detected thermal neutrons thus provides a quantitative measure of the shielding effectiveness against thermal neutrons. For thermal neutrons captured within the specified energy range, the shielding effectiveness is the captured neutron count ratio, defined as 

\[
R_{\text{det}} = \frac{N_{\text{shield}}}{N_{\text{bare}}}
\]

where \( N_{\text{shield}} \) is the number of measured neutron capture counts by EJ426 with shielding installed, and \( N_{\text{bare}} \) is the measured neutron capture counts by EJ426 without shielding (i.e., with the Am-Be source only). With this definition, the unshielded reference value is normalized to unity. 

\begin{figure}[htbp]
\centering
\includegraphics[width=.7\textwidth]{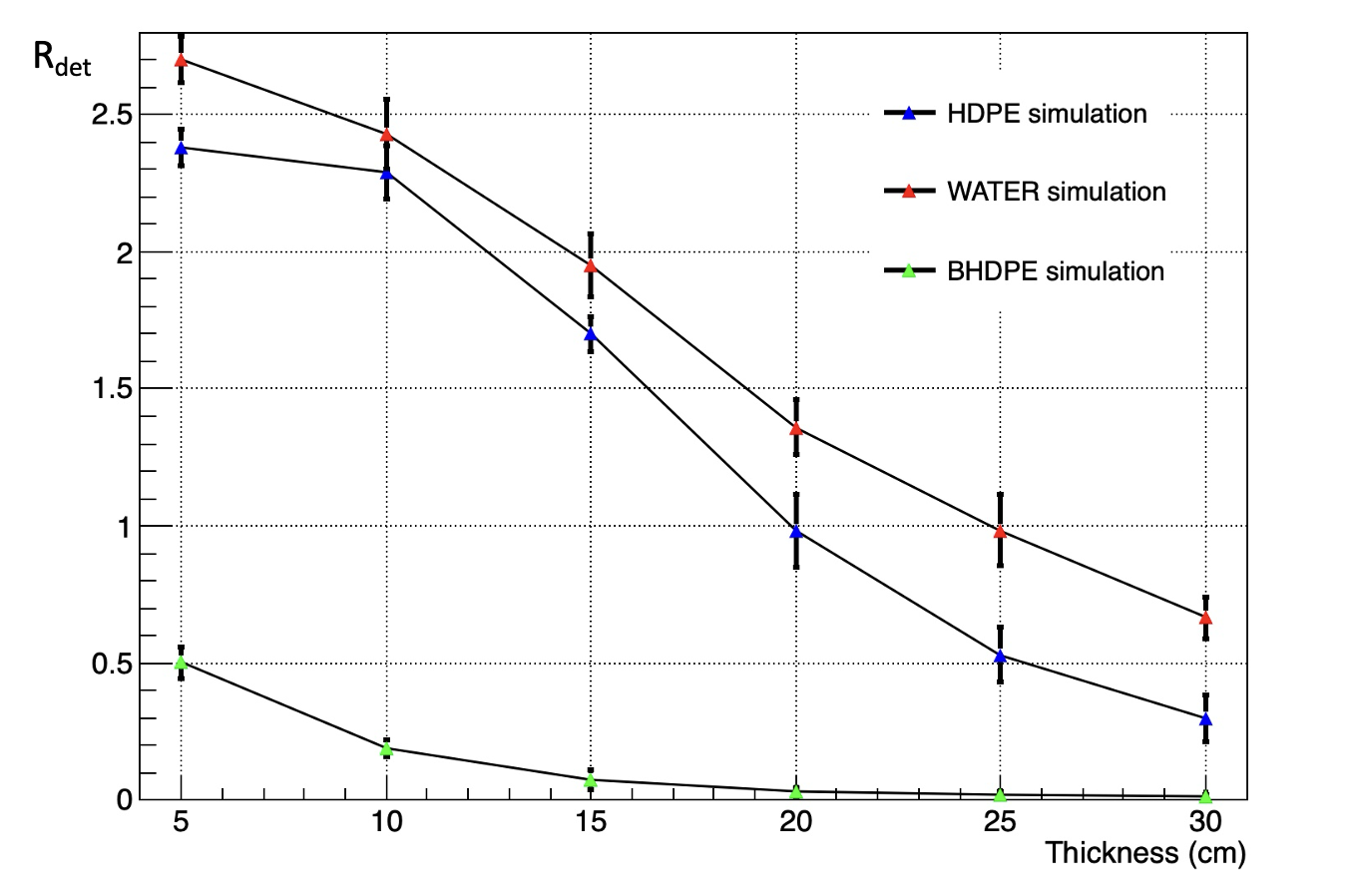}
\caption{Thermal neutron shielding efficiency simulated with the EJ426 model: captured neutron count ratio for different materials.
.\label{fig:simulation thermal}}
\end{figure}

Simulation results in figure~\ref{fig:simulation thermal} demonstrate that water and HDPE exhibit comparable shielding performance against thermal neutrons, while BHDPE shows significantly enhanced attenuation capabilities. These predictions can provide theoretical references for subsequent neutron shielding experiments.

\FloatBarrier

\section{Shielding Measurement for Single-Layer EJ426 detector}
\label{sec:results}

\subsection{Experimental setup}
\begin{figure}[htbp]
\centering
\includegraphics[width=.9\textwidth]{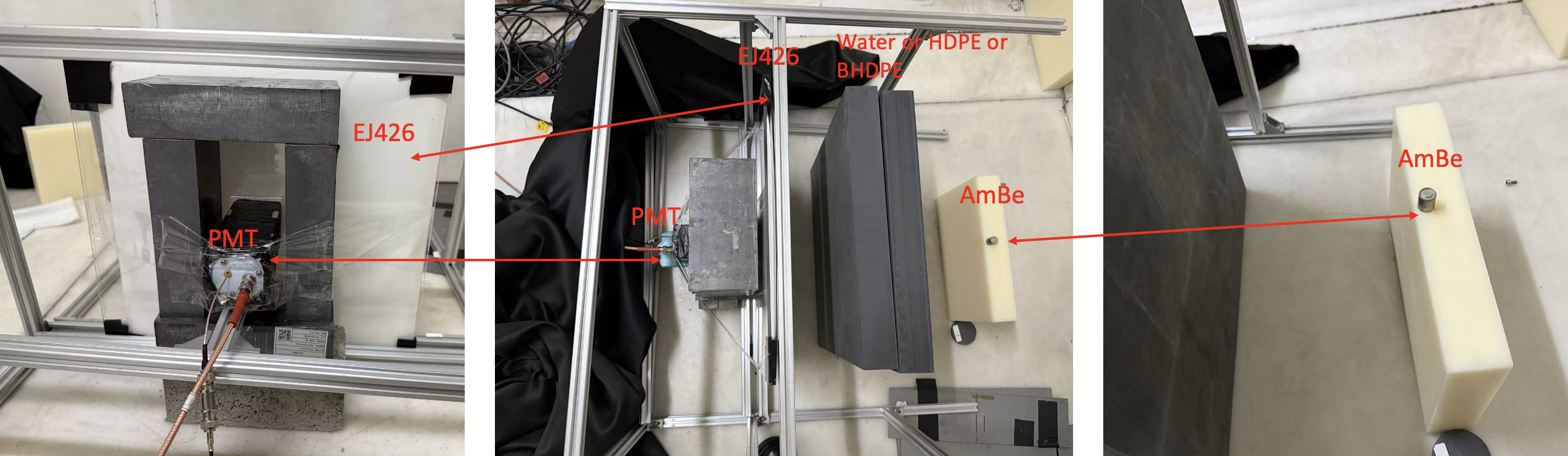}
\caption{Experimental setup: The configuration of EJ426 and PMT is shown on the left, the Am-Be source is placed on the right, the space in between is used to place various shielding materials.\label{fig:setup}}
\end{figure}

To test the neutron-shielding effectiveness of water, HDPE and BHDPE with a boron mass fraction of 40\%, we exposed a single-layer EJ426 scintillator to fast neutrons from an Am-Be source~\cite{ambe}. The water was stored in 3-mm-thick acrylic panels, while the HDPE and BHDPE were also shaped into panels. Each panel had a thickness of 5 cm and matches the dimensions of EJ426. The experimental setup is illustrated in the figure~\ref{fig:setup}, with EJ426 and PMT fixed on an aluminum frame and tightly secured using tape. The Am-Be source was also fixed at a position 31 cm away from EJ426, with the space in between to be used to place shielding materials of varying thicknesses.

The EJ426 scintillator, manufactured by ELJEN, is a square thin sheet measuring 420 mm × 420 mm × 0.5 mm, exhibiting a peak emission wavelength of 450 nm~\cite{eljen426}. Its central region is coupled to a PMT via HOTON SL610~\cite{hoton} silicone optical grease, the refractive index of which closely matches that of the scintillator, reducing the reflection loss of photons caused by air gaps~\cite{gierlik2006investigation}. All other sides not coupled to the PMT are wrapped with aluminum foil to enhance photon collection efficiency.

\begin{figure}[htbp]
\centering
\includegraphics[height=.28\textheight,width=.45\textwidth,keepaspectratio]{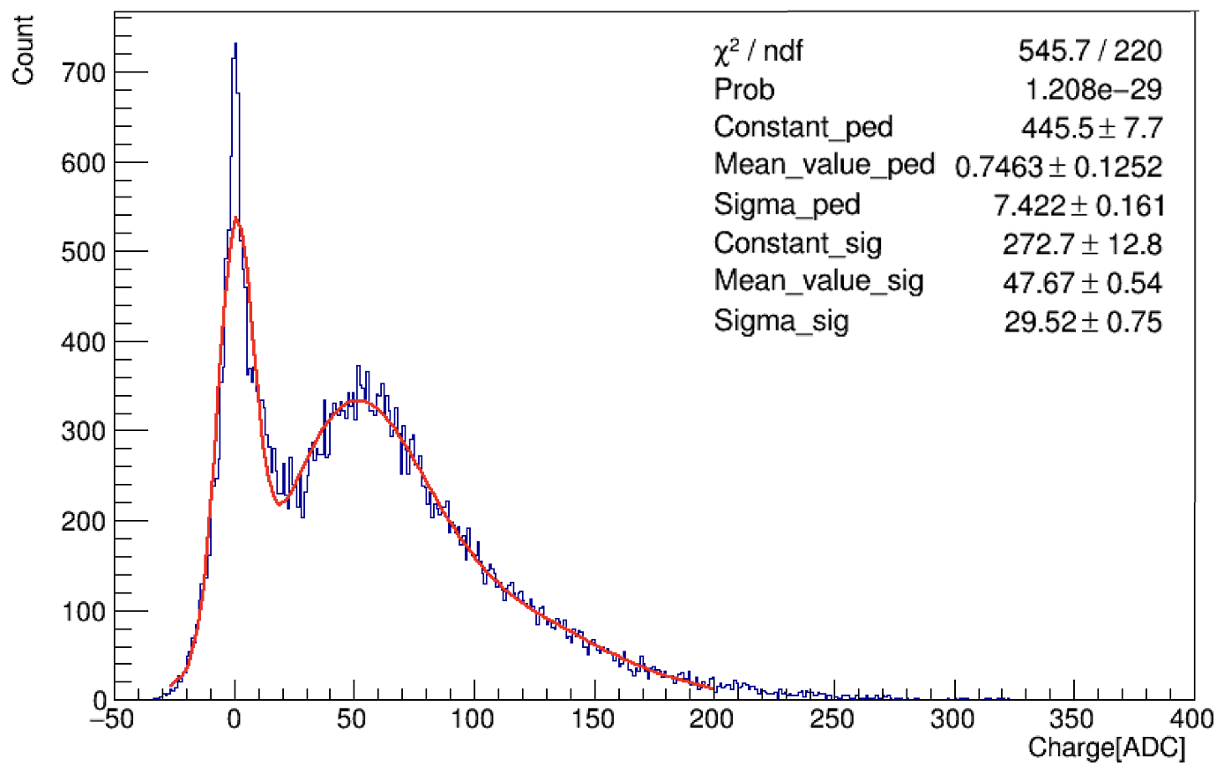}
\qquad
\includegraphics[height=.28\textheight,width=.45\textwidth,keepaspectratio]{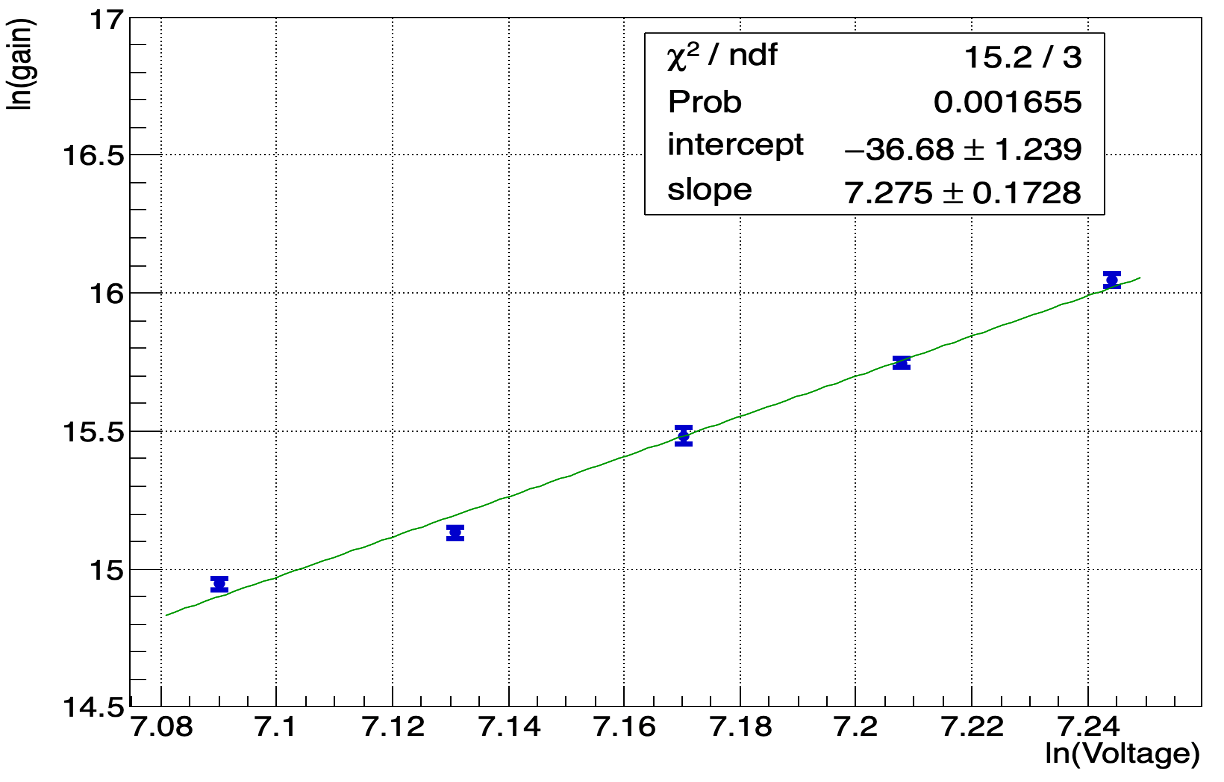}
\caption{Single-photoelectron spectrum of the XP3232 PMT with a double-Gaussian fit (left); gain curve of the XP3232 PMT measured at 1200, 1250, 1300, 1350, and 1400 V and shown as $\ln(G)$ versus $\ln(V)$ (right).\label{fig:2027}}
\end{figure}

The EJ426 scintillator was coupled with the XP3232 PMT produced by Hainan Zhanchuang Photonics Technology Corporation. The PMT features high photon detection efficiency, with a peak sensitivity wavelength of 420 nm, which matches the emission spectrum of the plastic scintillators. Using an LD laser as the light source, we measured the PMT gain at applied voltages of 1200, 1250, 1300, 1350, and 1400 V. The single-photoelectron charge spectra were fitted with a Double-Gaussian function, and the charge interval between the pedestal and the single-photoelectron peak was used to calculate the PMT gain. The gain curve was then obtained by fitting $\ln(G)$ as a function of $\ln(V)$, as shown in figure~\ref{fig:2027}. Based on the gain curve, we set the PMT gain to $3 \times 10^6$ and the operating voltage to 1250 V.

\begin{figure}[htbp]
\centering
\includegraphics[width=.4\textwidth]{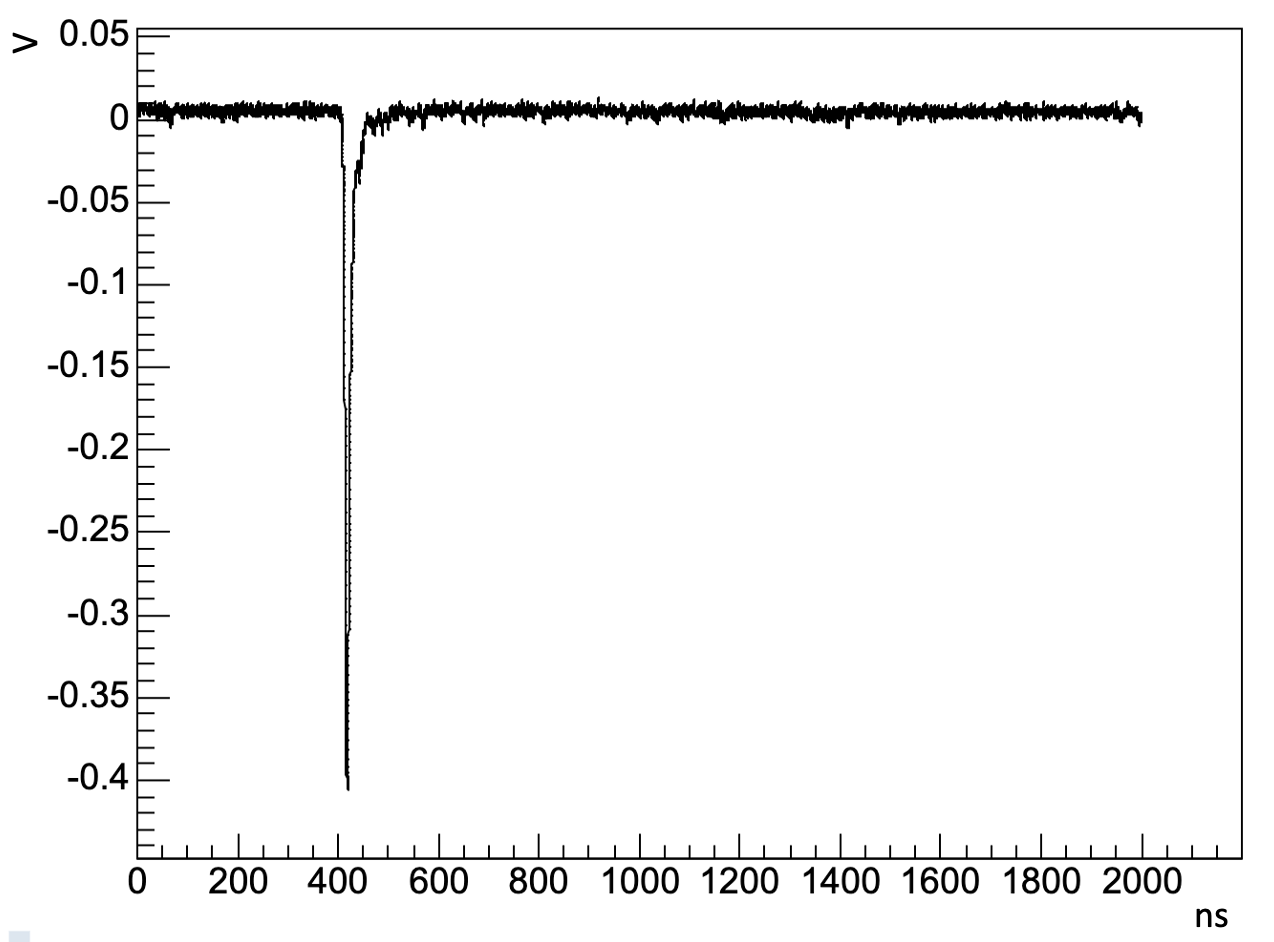}
\qquad
\includegraphics[width=.4\textwidth]{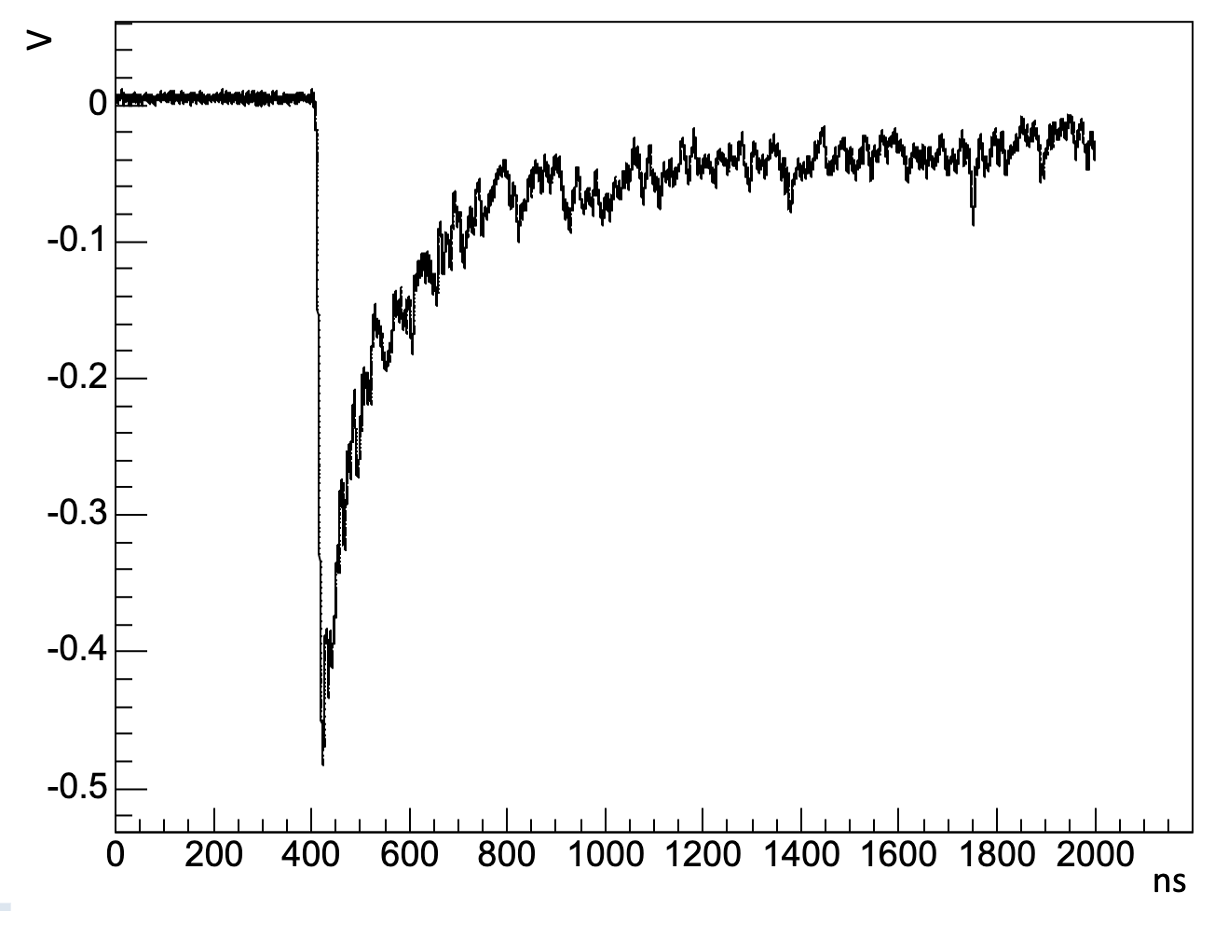}
\caption{The pulse shape of the signals produced by gamma (left) and the signal produced by thermal neutron (right).\label{fig:signal}}
\end{figure}

For the system composed of EJ426 and PMT, we used an oscilloscope with a time window of 2000 ns to capture the full pulse waveforms, as shown in figure~\ref{fig:signal}. The left figure shows the signal from the scintillator after it is excited by gamma rays emitted from the Am-Be source and produces photons, which are detected by the PMT and output to the oscilloscope. The right figure shows the thermal-neutron signal resulting from capture on lithium nuclei. It can be observed that the neutron signal possesses a longer falling-edge duration than the gamma signal. This difference in pulse shape constitutes the basis for pulse-shape discrimination (PSD) method. Definition of the PSD parameter is given in Equation~\eqref{eq:psd}~\cite{PSD_and_exploration}:

\begin{equation}
\mathrm{PSD} = 1 - \frac{Q_{\mathrm{Short}}}{Q_{\mathrm{Long}}}
\label{eq:psd}
\end{equation}

where \( Q_{\text{Long}} \) is the integrated charge over the entire waveform and \( Q_{\text{Short}} \) is the integrated charge over a short initial portion of the pulse.

\FloatBarrier

\subsection{Experimental results and discussion}
\begin{figure}[htbp]
\centering
\includegraphics[width=.4\textwidth]{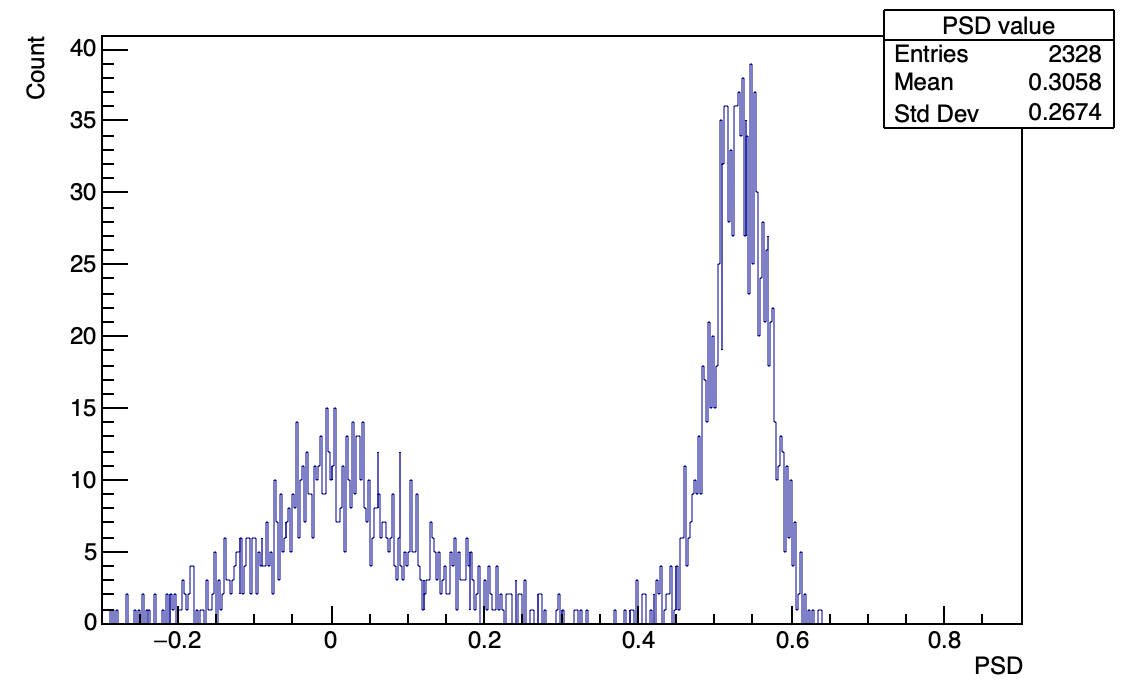}
\qquad
\includegraphics[width=.4\textwidth]{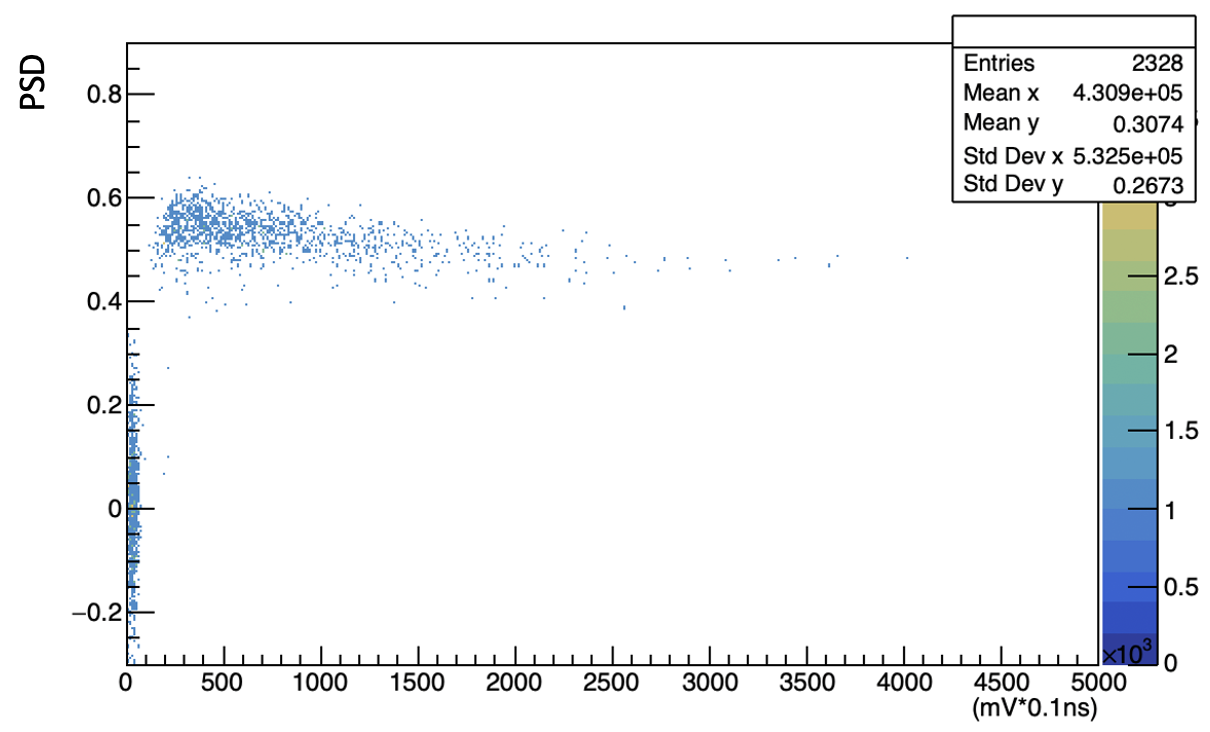}
\caption{An example of 1D (left) and 2D (right) PSD histograms of the gamma and neutron signals with a bare Am-Be source.\label{fig:psd}}
\end{figure}

\begin{figure}[htbp]
\centering
\begin{minipage}[t]{0.32\textwidth}
    \centering
    \includegraphics[width=\textwidth]{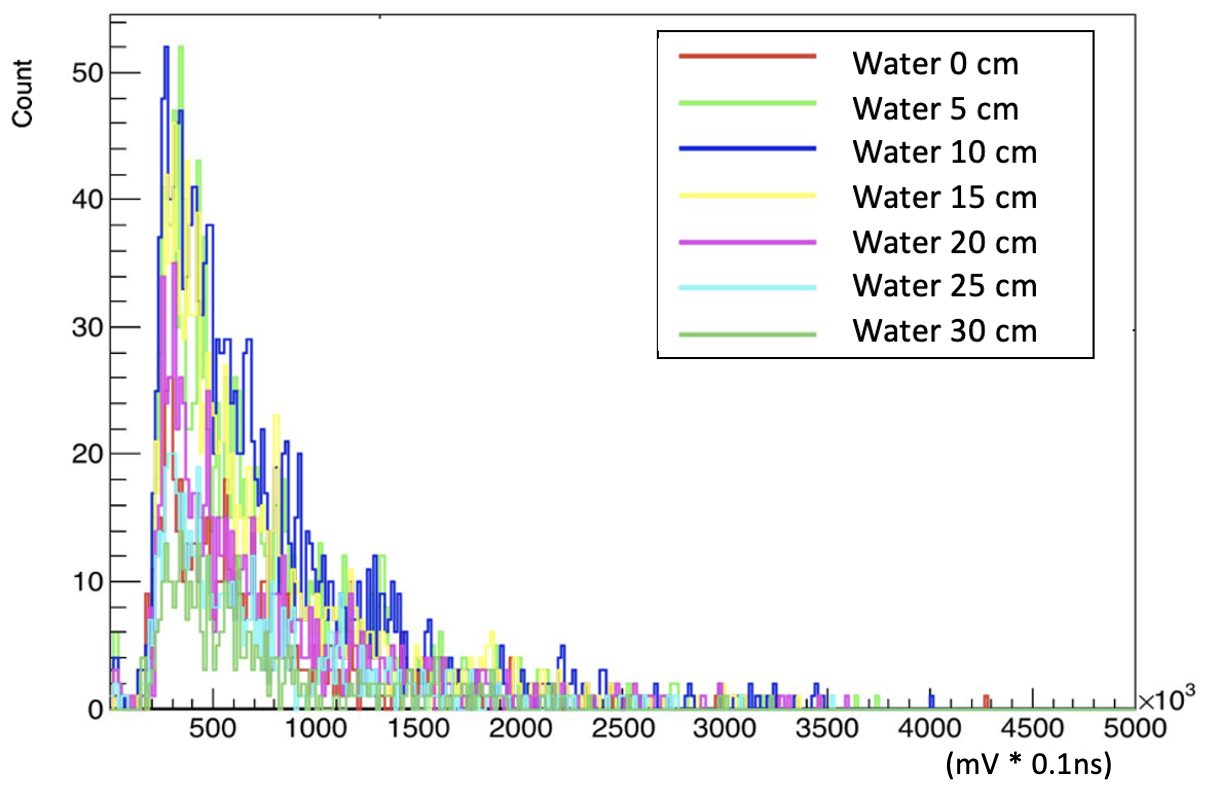}
\end{minipage}
\hfill
\begin{minipage}[t]{0.32\textwidth}
    \centering
    \includegraphics[width=\textwidth]{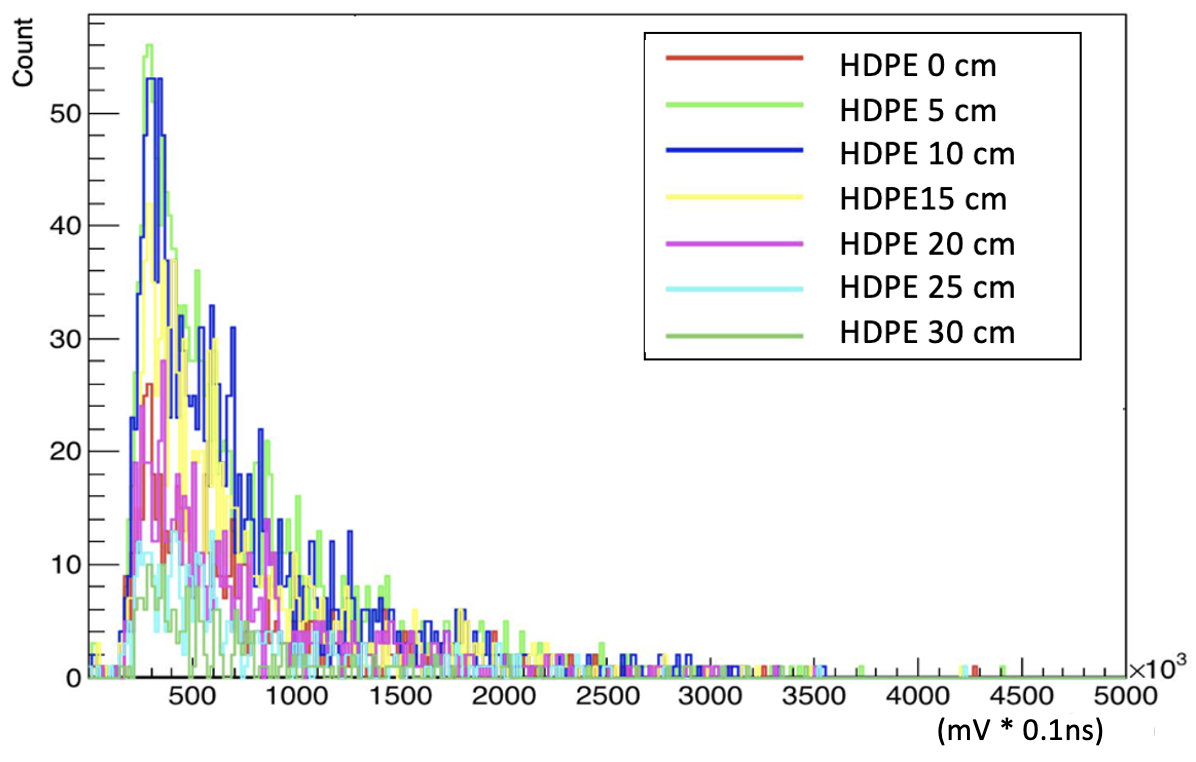}
\end{minipage}
\hfill
\begin{minipage}[t]{0.32\textwidth}
    \centering
    \includegraphics[width=\textwidth]{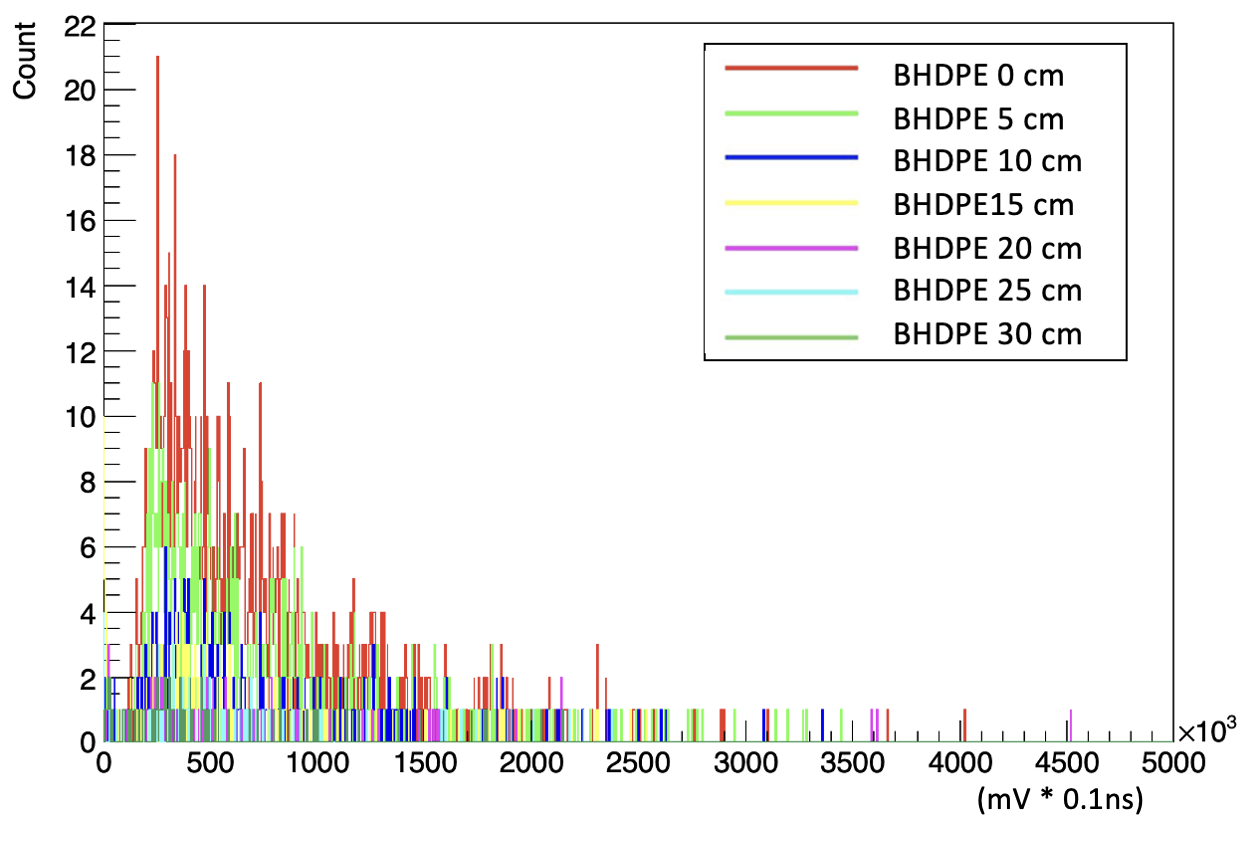}
\end{minipage}
\caption{Pulse integral distributions (background subtracted) of thermal neutrons for water (left), HDPE (center), and BHDPE (right), each measured with the Am-Be source exposure for 15 minutes.}
\label{fig:pulse integral}
\end{figure}

Using Equation (\ref{eq:psd}), we calculate the PSD value for each waveform. The figure of merit (FOM), defined as \(\mathrm{FOM} = (S_n - S_y)/(\mathrm{FWHM}_n + \mathrm{FWHM}_y)\)~\cite{jiang2022study}, quantifies the separation between neutron and gamma peaks in the one-dimensional PSD spectrum, where \(S_n\) and \(S_y\) are the peak positions and \(\mathrm{FWHM}_n\) and \(\mathrm{FWHM}_y\) are the full widths at half maximum. A larger FOM indicates better discrimination capability. Examples of 1D and 2D PSD distributions obtained with a standalone Am-Be source are provided in figure~\ref{fig:psd}. In the 1D histogram, the left peak corresponds to gamma events while the larger right peak represents neutron events (neutron signals exhibit higher PSD values due to their longer decay tails). The 2D histogram shows a horizontal band in the upper region for neutrons and a vertical band in the lower section for gammas, demonstrating clear separation between the two signals. Based on the PSD distribution and the FOM value, a threshold of \(\mathrm{PSD} > 0.4\) was selected to identify neutron events for subsequent analysis. In figure~\ref{fig:pulse integral}, pulse integral distributions were obtained for water, HDPE, and BHDPE shielding at thicknesses of 5, 10, 15, 20, 25, and 30 cm, revealing that thermal neutron events mainly cluster around 50,000 mV·ns.

\begin{figure}[htbp]
    \centering
    \begin{subfigure}[b]{0.45\textwidth}
        \includegraphics[width=\textwidth]{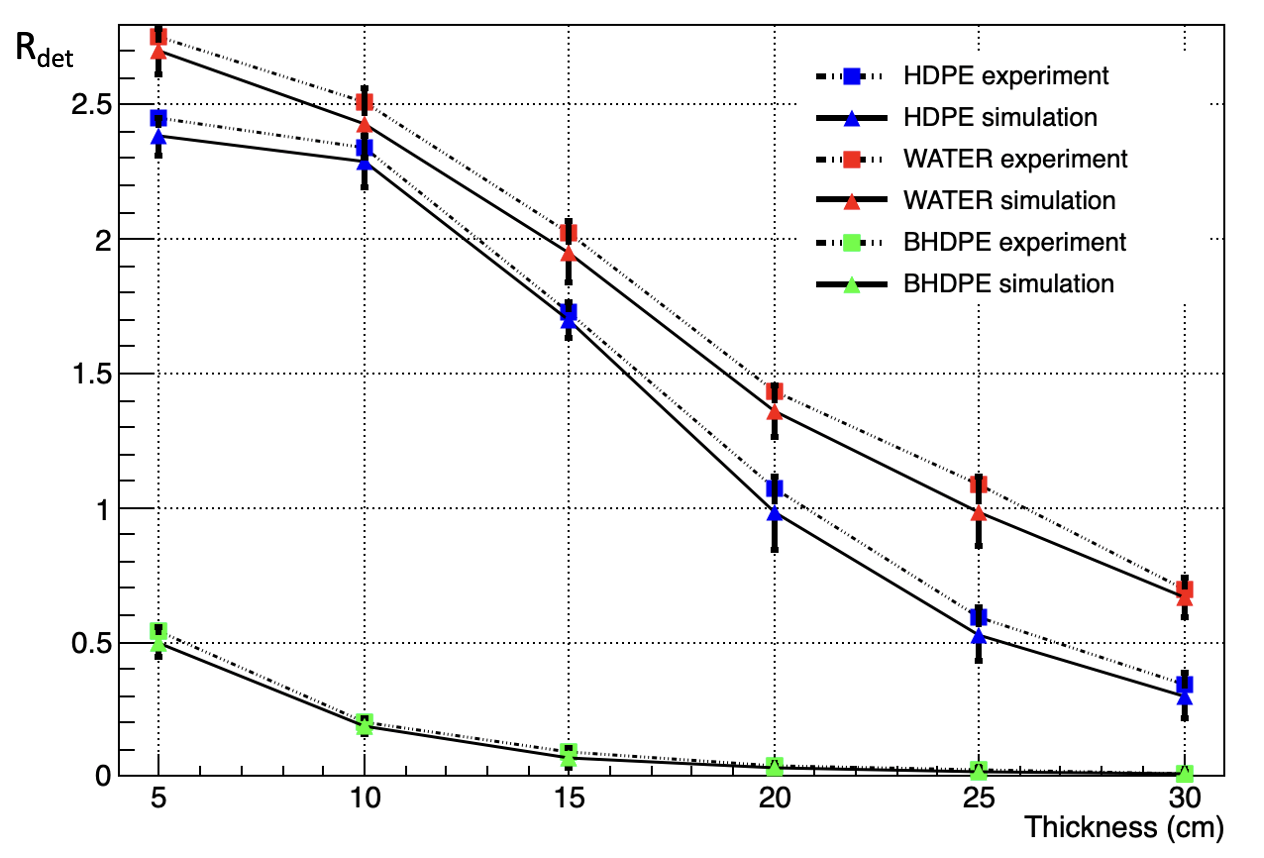}
        \subcaption{Experimental captured neutron count ratio with simulation reference}
        \label{fig:i}
    \end{subfigure}
    \hfill
    \begin{subfigure}[b]{0.45\textwidth}
        \includegraphics[width=\textwidth]{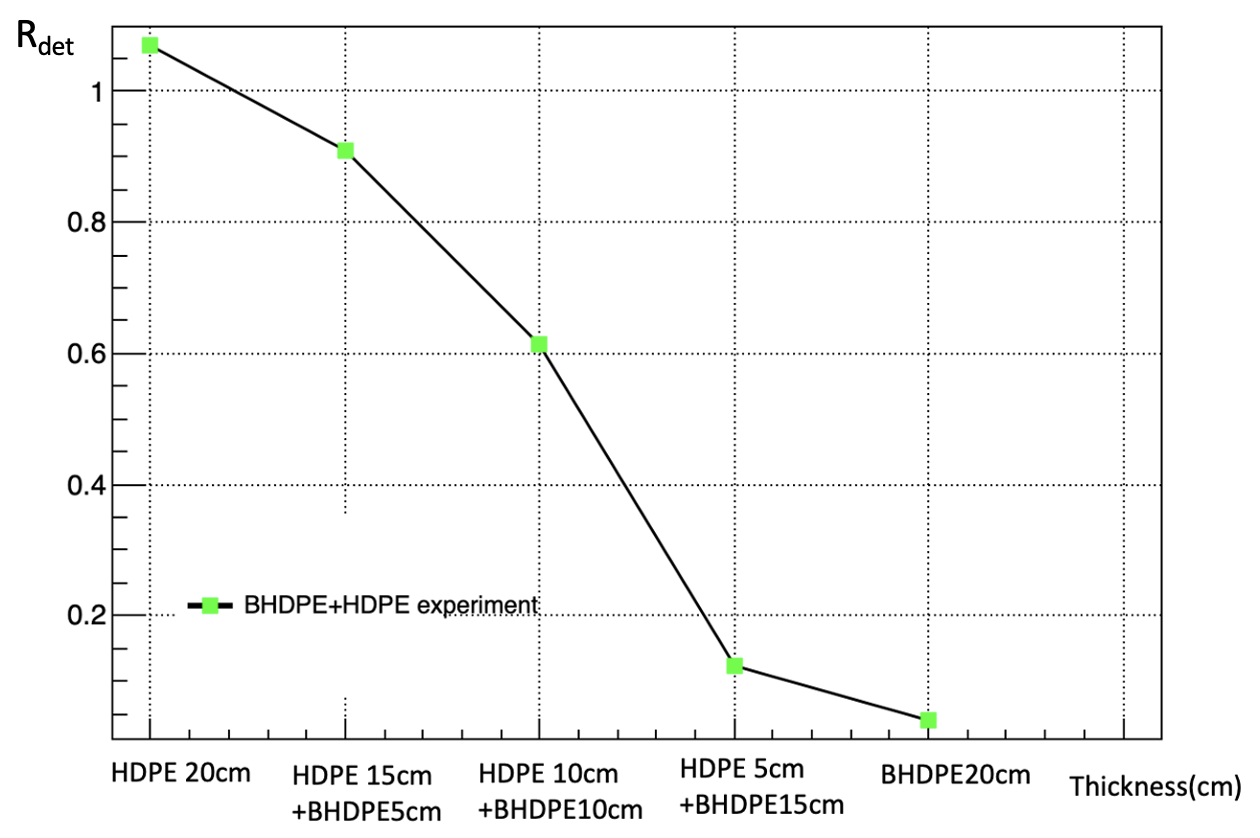}
        \subcaption{Experimental performance of composite shielding: HDPE/BHDPE layered configurations}
        \label{fig:combine}
    \end{subfigure}
    \caption{Experimental results: (a) captured neutron count ratio with simulation reference; (b) performance of composite shielding with HDPE/BHDPE layered configurations.}
    \label{fig:combined}
\end{figure}

By integrating the full pulse integral range in figure~\ref{fig:pulse integral}, the thermal neutron capture counts were obtained and are presented in table~\ref{tab:neutron counts}. All counts reported in table~\ref{tab:neutron counts} have been background-subtracted. Using the bare Am-Be source condition as the baseline, we normalized the neutron counts as \(R_{\mathrm{det}}\) and compared them with simulation results. As shown in figure~\ref{fig:combined}~(\subref{fig:i}), experimental data agrees well with the simulation result within the uncertainty bands. This good agreement validates the reliability of the subsequent simulations for the ALARM detector. For water and HDPE, adding shielding initially increases thermal neutron counts significantly, as fast neutrons are moderated into thermal neutrons via scattering interactions with the shielding material~\cite{introduction_shielding}. Between 5–10 cm thicknesses, a gradual reduction occurs as fast neutrons lose energy through elastic scattering, while thermalization is not yet fully achieved. From 10-20 cm, rapid attenuation dominates with efficient thermalization of most neutrons. Beyond 20 cm, the decline rate slows as only a few portion of neutrons remain unmoderated. HDPE offers better thermal neutron attenuation than water at equivalent thicknesses, with an improvement ranging from 9\% to 48\% across various thicknesses, while the maximum enhancement of 50\% appears at the thickness of 30 cm. BHDPE, however, achieves a dramatic reduction in thermal neutron counts compared to both HDPE and water due to the addition of neutron-absorbing boron. Its shielding efficiency exceeds 95\% at just 20 cm thickness.

\begin{table}[htbp]
\centering
\caption{Thermal neutron capture counts under varied shielding configurations.\label{tab:neutron counts}}
\smallskip
\begin{tabular}{|c|c|c|c|c|c|c|c|}
\hline
  & Am-Be only & 5 cm & 10 cm & 15 cm & 20 cm & 25 cm & 30 cm\\
\hline
water & 638 & 1759 & 1604 & 1294 & 917 & 698 & 452\\
\hline
HDPE & 638 & 1562 & 1492 & 1103 & 680 & 378 & 219\\
\hline
BHDPE & 834 & 450 & 169 & 76 & 33 & 21 & 12\\
\hline
\end{tabular}
\end{table}


We further evaluated composite shielding configurations combining HDPE and BHDPE layers, contrasting their performance against pure HDPE and BHDPE, as shown in figure~\ref{fig:combined}~(\subref{fig:combine}). Maintaining a constant total thickness of 20 cm, thermal neutron capture rates exhibit rapid reduction as BHDPE fraction increases. Remarkably, the neutron capture rates obtained with the HDPE(5 cm)+BHDPE(15 cm) configuration are no more than 10\% higher than those with pure 20 cm BHDPE.



\FloatBarrier

\section{Shielding Simulation for ALARM Detector}
\label{sec:alarmresults}
To provide a more accurate estimate of the shielding efficiency of different materials under the neutron environment of the Taishan experimental hall, a Geant4-based model of the ALARM detector was developed. The model includes the geometry of the detector, the properties of materials used, and the characteristics of the neutron source.

\subsection{ALARM physics model}
\begin{figure}[htbp]
\centering
\includegraphics[width=.7\textwidth]{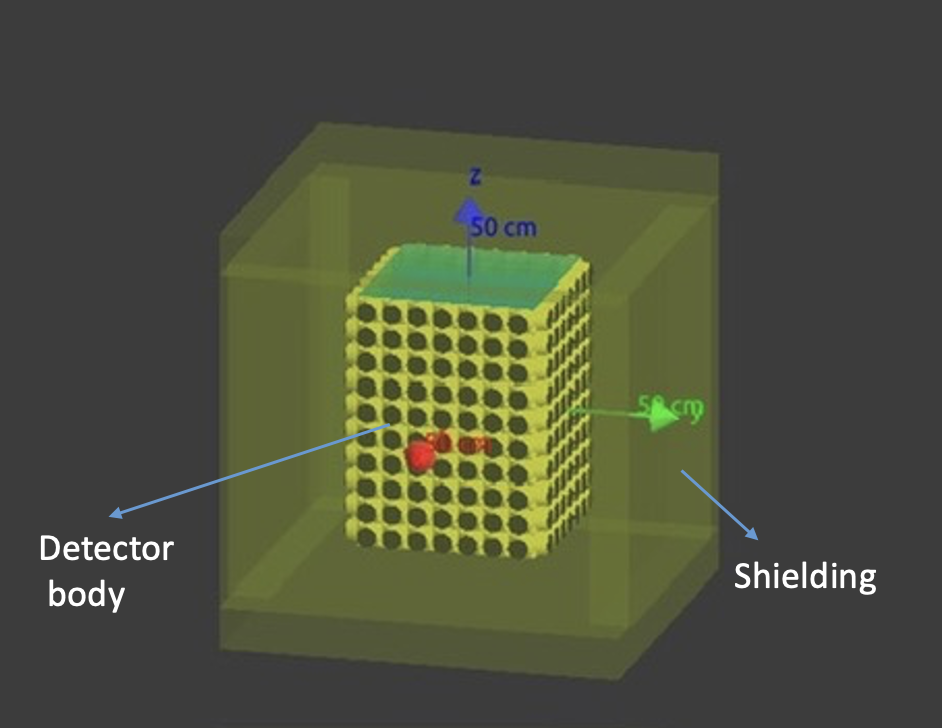}
\caption{General 3D view of the ALARM detector model geometry.\label{fig:detector_setup}}
\end{figure}

In the Monte Carlo simulation, the ALARM neutron detector is modeled as a 7×7×10 array of EJ200 scintillator cubes, interleaved with 11 layers of EJ426 scintillator sheets. Each outermost EJ200 cube on every layer is coupled to a PMT, totaling 280 PMTs. The energy distribution of the neutron source is based on the measured neutron spectrum from the Taishan environment, imported into Geant4 using the General Particle Source (GPS) method. Neutrons are emitted uniformly from a spherical surface with a radius of 150 cm toward the center of the detector. The detector is fully enclosed by shielding layers on all six sides, each side having the same thickness and composed of either water, HDPE, or BHDPE. The geometry of the simulation setup is illustrated in figure~\ref{fig:detector_setup}. The program records the same physics data in each run as the EJ426 Monte Carlo simulation does.

\FloatBarrier

\subsection{ALARM analysis results}
Analysis of the ALARM detector simulation provides a reliable basis for estimating the shielding performance of water, HDPE, and BHDPE at different thicknesses against fast and thermal neutrons under real experimental-hall conditions.

\begin{figure}[htbp]
    \centering
    \begin{subfigure}[b]{0.45\textwidth}
        \includegraphics[width=\textwidth]{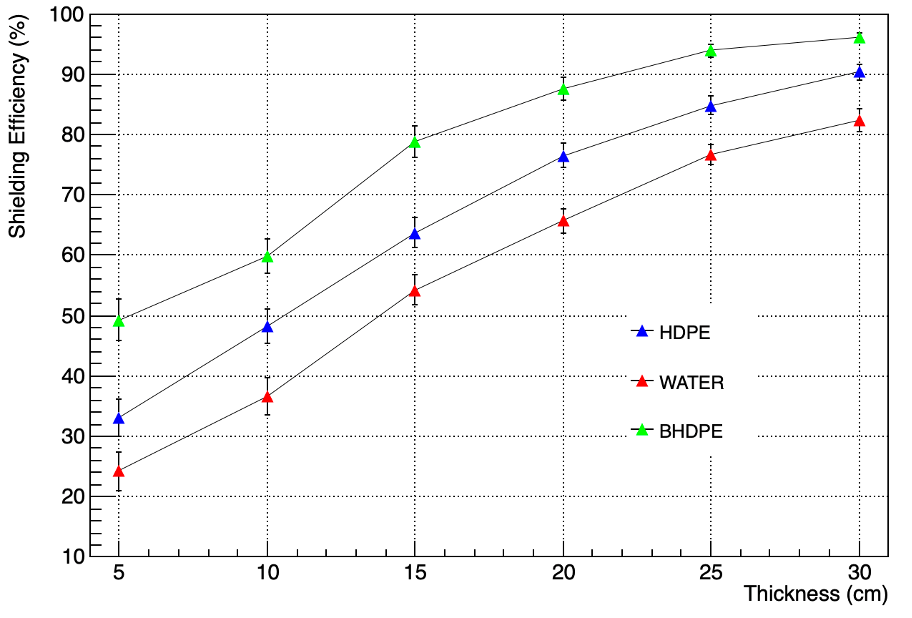}
        \subcaption{Fast neutron shielding efficiency}
        \label{fig:detector_fastneutron}
    \end{subfigure}
    \hfill
    \begin{subfigure}[b]{0.45\textwidth}
        \includegraphics[width=\textwidth]{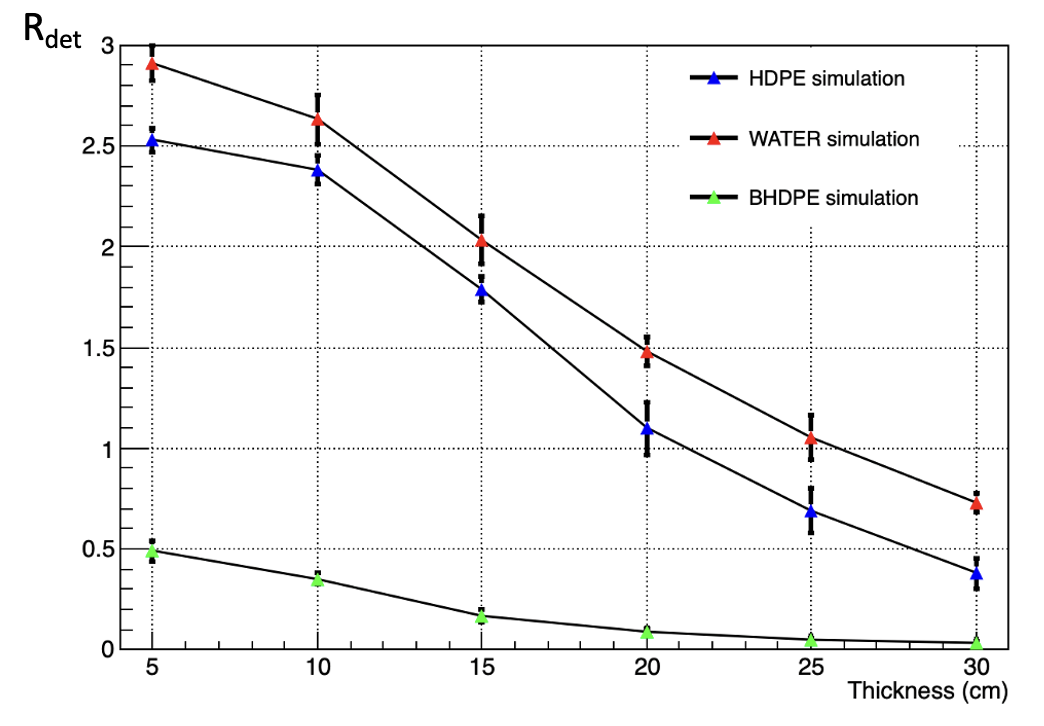}
        \subcaption{Thermal neutron shielding efficiency}
        \label{fig:detector_thermalneutron}
    \end{subfigure}
    \caption{Simulation results with ALARM detector model for different materials: (a) Fast neutron shielding efficiency; (b) Thermal neutron shielding efficiency}
    \label{fig:combined_alarm}
\end{figure}

Figure~\ref{fig:combined_alarm}~(\subref{fig:detector_fastneutron}) presents the shielding efficiency against the fast neutrons. It can be observed that for all three shielding materials, the efficiency increases with greater thickness, while the rate of improvement gradually slows. At equivalent thicknesses, HDPE consistently achieves approximately 10\% higher shielding efficiency than water, whereas BHDPE exceeds pure HDPE by an average of more than 10\%. At a thickness of 30 cm, BHDPE still attains a shielding efficiency above 90\%.

Compared with the results from the single‑layer EJ426 simulation, the detector simulation—which uses the measured neutron energy spectrum of the Taishan environment shared by the Taishan Neutrino Observatory (TAO) instead of the Am-Be spectrum, gives slightly higher fast-neutron shielding efficiencies at a thickness of 5 cm for all three materials, whereas the efficiencies are about 5\%--15\% lower in the 10--30 cm thickness range.


For the captured thermal neutron counts, the variation is still quantified by the captured neutron count ratio. As shown in figure~\ref{fig:combined_alarm}~(\subref{fig:detector_thermalneutron}), the capture rates for HDPE and water follow similar trends, both decreasing steadily as the shielding thickness increases. At equivalent thicknesses, the rate for HDPE is approximately 10–35\% lower than that for water. BHDPE shows a significantly enhanced shielding effect, reducing the thermal neutron count to less than 1/25 of the unshielded case at 30 cm. 

Compared with the single‑layer EJ426 simulation, the ALARM detector simulation yields slightly lower thermal‑neutron shielding rates for the three materials due to the different detector geometry and neutron energy spectrum, but the difference becomes negligible at a thickness of 30 cm, where both simulations give a rate above 95\% for BHDPE.


\FloatBarrier

\section{Conclusion}
In this study, a detector system comprising a single-layer EJ426 scintillator coupled to an XP3232 PMT was first employed to evaluate the shielding efficiency of water, HDPE, and BHDPE at various thicknesses against neutrons emitted from an Am–Be source. In terms of thermal-neutron shielding efficiency, as shielding thickness increases from 5 cm to 30 cm, the measured thermal-neutron capture counts for both water and HDPE initially rose and then gradually declined compared to the unshielded (Am-Be-only) case. HDPE outperformed water by about 9\%–48\% at equivalent thicknesses. BHDPE, containing $^{10}\text{B}$ nuclei with high neutron absorption cross-sections, displayed markedly higher shielding efficiency compared to the other two materials. When HDPE was combined with BHDPE, the thermal-neutron capture counts dropped progressively with increasing BHDPE fraction. This offers a viable option when seeking effective shielding at relatively lower cost. Monte Carlo simulations were also used to predict the neutron shielding performance of the single-layer EJ426 setup, and the simulation results agreed well with the experimental data.

Subsequently, we simulated the neutron-shielding performance of the ALARM detector under realistic experimental conditions. The results show that the shielding effectiveness of the ALARM detector is comparable to that of the single‑layer EJ426 setup, both for fast neutrons and for thermal neutrons. Among the three materials studied, BHDPE provides the best shielding performance, achieving an efficiency for both fast and thermal neutrons above 95\% at a thickness of 30 cm. Consequently, 30‑cm‑thick BHDPE will be adopted for the ALARM experiment. These simulation results for the ALARM detector can serve as a useful reference for detectors with similar neutron shielding requirements.

\appendix

\acknowledgments

This research was supported by the Fundamental Research Funds for the Central Universities (Grant No. 24qnpy109) and the China Postdoctoral Science Foundation (Grant No. 2025M783462) and the National Natural Science Foundation of China (Grant No. 12075087). We thank the School of Physics and Institut Franco-Chinois de l’Énergie Nucléaire at Sun Yat-sen University. We are also grateful to Prof. Yuehuan Wei, Prof. Bo Mei, and Prof. Tao Xiong for their support.





\bibliographystyle{JHEP}
\bibliography{biblio.bib}

\end{document}